\documentclass[aps,prb,twocolumn]{revtex4-1}
\usepackage{amsmath}
\usepackage{amsfonts}
\usepackage{amssymb}
\usepackage{graphicx}
\usepackage[colorlinks=true,linktocpage=true,breaklinks=true]{hyperref}

\begin{document}

\newcommand{\nnm}{\nonumber \\}
\newcommand{\ls}{\lambda_{SO}}
\newcommand{\lr}{\lambda_R}
\newcommand{\ez}{E_Z}
\newcommand{\ec}{E_C}
\newcommand{\pc}{persistent current}

\title{Signature of topological transition in persistent current in a Dirac Ring}
\author{S. Ghosh}
\email{sumit.ghosh@kaust.edu.sa}
\author{A. Manchon}
\email{aurelien.manchon@kaust.edu.sa}
\affiliation{Physical Science and Engineering Division (PSE),\\
King Abdullah University of Science and Technology (KAUST),\\
Thuwal 23955-6900, Kingdom of Saudi Arabia}

\begin{abstract}
We study the \pc\ in a one dimensional Dirac ring and show that the change of spin current with respect to an applied perpendicular electric field can be used to identify the topological phases. We further study the effect of Rashba spin orbit coupling and show that the Aharonov-Casher phase appearing due to Rashba spin orbit coupling vanishes in topologically nontrivial regime and thus can identify the topological phases. This Aharonov-Casher phase causes a finite spin-valley current in presence of valley mixing perturbation and is thus useful to detect the topological phases even in presence of such impurity. 
\end{abstract}

\maketitle

\section{Introduction}
When a normal metal ring with dimensions comparable to the phase coherence length of the system is subjected to an Aharonov-Bohm flux, it can support a flux-periodic persistent current \cite{Buttiker1983}. This phenomenon has led to quite extensive theoretical \cite{Cheung1988, Loss1992, Chakraborty1995} and experimental \cite{Levy1990, Mailly1993, Jariwala2001, Bleszynski-Jayich2009} studies. The current is linear in flux, which is a result of parabolic dispersion relation of the nonrelativistic particles. During the last decade a new class of two dimensional materials (e.g. graphene and its siblings and HgTe quantum well) has been discovered where electrons follow a non-quadratic dispersion. Among other phenomena, this discovery led to a new boost in the studies of Aharonov-Bohm effect and persistent currents, from both theoretical \cite{Recher2007, Zarenia2010, Schelter2012} and experimental \cite{Russo2008, Peng2010} perspectives. These materials can mimic the behaviour of a relativistic particle and thus provides a unique opportunity to study different relativistic phenomena and make quantum rings an ideal testing ground \cite{Fertig2010, Yannouleas2014}.  Apart from their dispersion relation, another feature that makes them exotic is their ability to undergo a topological transition. The topological transition is triggered by a strong spin orbit coupling and manifests itself as zero energy edge modes in a one dimensional system \cite{Kane2005a}. Such edge states have already been predicted in a wide topological insulator ring \cite{Michetti2011}. However the edge states have a finite depth of penetration and hence disappear for a very narrow system \cite{Zhou2008}. In such cases it is difficult to detect a topological phase transition. In this paper we propose a way to detect topological transition by studying the persistent current in a one dimensional ring. Moreover, in the present of Rashba spin-orbit coupling, we show that it is possible to detect the topological transition by observing the persistent spin current as well as the phase-shift due to Aharonov-Casher even in the presence of valley mixing impurities.
For our study we choose a honeycomb lattice in presence of spin orbit coupling. We focus on the buckled lattices like silicene or germanene because here one can easily tune the topological phases by an applied electric field \cite{Ezawa2012a, Drummond2012}, which gives us additional flexibility.

\section{Energy spectrum and Persistent current in a Si/Ge ring}

The effective Hamiltonian around the $K$ and $K'$ points for a buckled honeycomb lattice is given by \cite{Ezawa2012a, Ezawa2012} .
\begin{eqnarray}
H &=& \hbar v_F (\eta k_x \tau_x + k_y \tau_y) - \ell \ez \tau_z + \eta \tau_z h_0 \nnm
h_0 &=& \ls \sigma_z + a \lr (k_y \sigma_x - k_x \sigma_y)
\label{H}
\end{eqnarray}
where $\tau$ and $\sigma$ are Pauli matrices corresponding to the valley and spin. $\eta=\pm1$ is the valley index, $a$ is the interatomic distance, $\ell$ is the buckling parameter, $v_F$ is Fermi velocity, $\ez$ is an uniform electric field applied perpendicular to the plane, $\ls$ is the spin orbit coupling and $\lr$ is the Rashba parameter due to second nearest neighbour. The parameter values for different materials are given in Table~\ref{param}.
\begin{table}[h]
\caption{Material's parameters for graphene(Gr), silicene(Si), germanene(Ge) and stanene(Sn) \cite{Liu2011,Ezawa2015a}. }
{\begin{tabular}{|c|c|c|c|c|c|c|}
\hline
Atom & $a$ & $\ell $ & $\lambda_{SO} $ & $ \lambda_{R}$ & $v_F$ & $E_C$ \\
 & $\rm (\AA)$ & $\rm (\AA)$ & $\rm (meV)$ & $ \rm (meV)$  & $\rm 10^5 m/s$ & $ \rm (meV/\AA)$ \\ \colrule
Gr & 2.46 & 0.00 &$10^{-3}$& 0.0 & 9.8 & $\infty$ \\ \hline
Si & 3.86 & 0.23 &  3.9  & 0.7 & 5.5 & 17  \\ \hline
Ge & 4.02 & 0.33 & 43.0  &10.7 & 4.6 &130.3\\ \hline 
Sn & 4.70 & 0.40 & 100   & 9.5 & 4.9 &250  \\ \hline 
\end{tabular}}
\label{param}
\end{table}

\begin{figure}[h]
\centering
\includegraphics[width=0.4\textwidth]{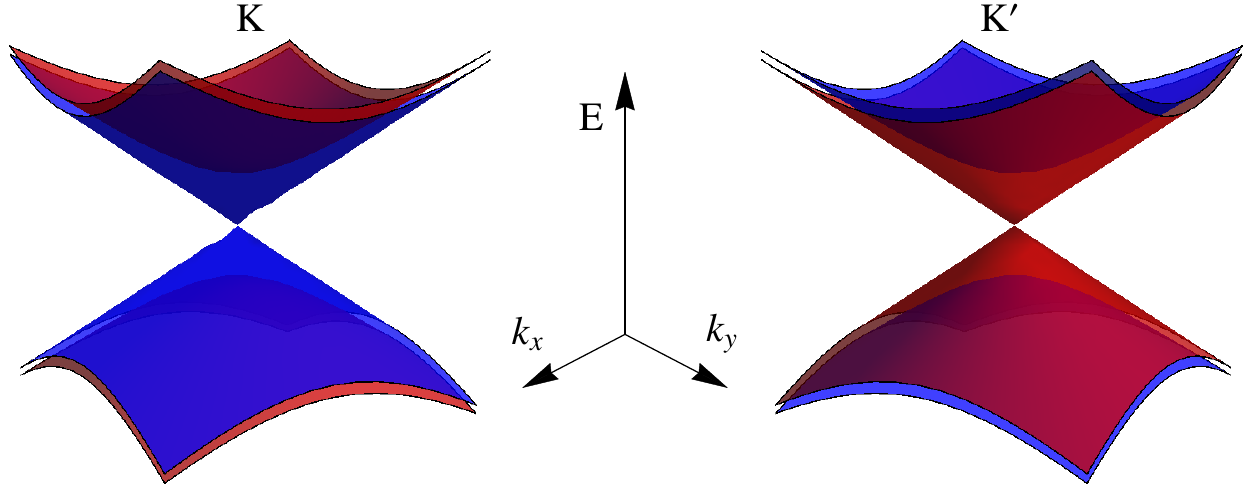}
\caption{Band structure from the Hamiltonian (\ref{H}) at $\ez=\ec$. Blue and red surfaces corresponds spin up and spin down.}
\label{figcone}
\end{figure}

For $\ez = \ec = \ls/\ell$, the Hamiltonian has two zeros corresponding to $\eta=1(K)$ and $\eta=-1(K')$ (Fig.~\ref{figcone}). However at these two valleys ($K$ and $K'$) the zero energy modes are formed by the opposite spin species (Fig.~\ref{figcone}).

Away from the critical field ($\ez \neq \ec$) the band gap can be controlled by the applied electric field \cite{Drummond2012}, which in turn also controls the ground state spin. We exploit this property to identify the topological phases in a ring geometry.
The Hamiltonian for a ring geometry can easily be obtained from (\ref{H}) in two steps : (1) replacing $k_x$ and $k_y$ by  $-i\frac{\partial}{\partial x}$ and $-i\frac{\partial}{\partial y}$ and (2) transforming the differential operator from cartesian ($x,y$) to polar ($r,\theta$) coordinates. 
\begin{eqnarray}
H_{r,\theta} &=& \hbar v_F (\tau_r^\eta p_r + \tau_\theta^\eta p_\theta) - \ell \ez \tau_z + \eta \tau_z h_1 \nnm
h_1 &=& \ls \sigma_z + a \lr (\sigma_r p_r + \sigma_\theta p_\theta) \nnm
p_r &=& -i\frac{\partial}{\partial r}, \hspace{1cm} p_\theta = -i\frac{1}{r}\frac{\partial}{\partial \theta} \nnm
\tau_r^\eta &=& \eta \cos(\theta) \tau_x + \sin(\theta) \tau_y \nnm
\tau_\theta^\eta &=& \cos(\theta) \tau_y - \eta \sin(\theta) \tau_x \nnm
\sigma_r &=& \sin(\theta) \sigma_x - \cos(\theta) \sigma_y \nnm
\sigma_\theta &=& \cos(\theta) \sigma_x + \sin(\theta) \sigma_y
\label{Hr}
\end{eqnarray}

\begin{figure}[h]
\centering
\includegraphics[width=0.3\textwidth]{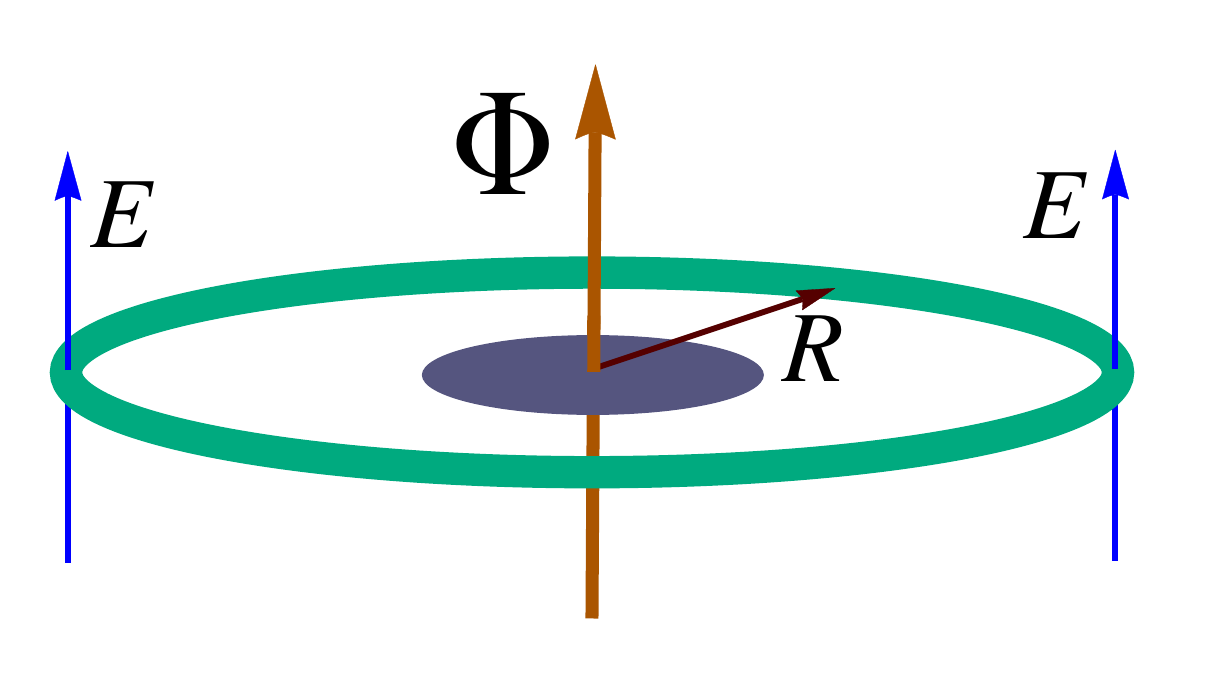}
\caption{A circular ring with radius R, threaded by an AB flux $\Phi$ and subjected to perpendicular electric field $E$.}
\label{figring}
\end{figure}
For a narrow ring (Fig.~\ref{figring}) an effective one dimensional Hamiltonian can be obtained by replacing $\frac{1}{r} \rightarrow \frac{1}{R}$ and $\frac{\partial}{\partial r} \rightarrow -\frac{1}{2R}$ \cite{Meijer2002}. Such one dimensional Hamiltonian has extensively used for graphene \cite{Zarenia2010}, Rashba \cite{Wang2011} and Dirac \cite{Ghosh2014} rings. In this paper we choose a ring radius $R$=50nm \cite{Zarenia2010}. One should be very careful at this stage. By proceeding with $\frac{\partial}{\partial r} \rightarrow 0$, one can encounter a nonhermitian Hamiltonian if the circumference of the ring is smaller than the Compton length (characterized by the gap) of the particle \cite{Ghosh2013} resulting in a divergence in persistent current. 
Next we will consider an Aharonov-Bohm flux threading the ring. Under symmetric gauge ($\vec{A}= \frac{\Phi}{2 \pi R} \hat{\theta}$) the effect can easily be incorporated by the substitution $-i\frac{\partial}{\partial \theta} \rightarrow -i\frac{\partial}{\partial \theta} + \frac{\Phi}{\Phi_0}$, where $\Phi_0$ is the flux quantum. 

\begin{figure}[h]
\centering
\includegraphics[width=0.4\textwidth]{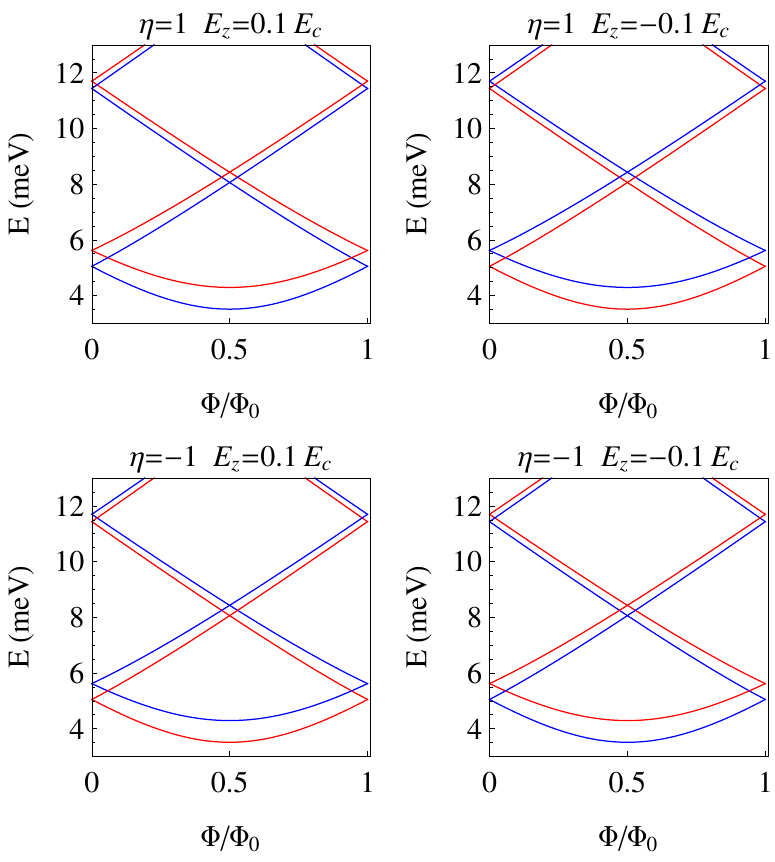}
\caption{Band structure of silicene (Table\ref{param}) ring ($R=50\rm nm$) at $\ez=\pm 0.1 \ec$ and $\lr=0$ for . Blue and red lines corresponds spin up and spin down.}
\label{figsiband}
\end{figure}

We can diagonalize the Hamiltonian (\ref{Hr}) with the basis $(\psi_\uparrow^A, \psi_\uparrow^B, \psi_\downarrow^A, \psi_\downarrow^B)^T$. 
For simplicity we first ignore Rashba coupling. From Table~\ref{param} we see that $\lr$ is one order of magnitude smaller than $\ls$ and is coupled to the momentum. So long we are close to the $K,K'$ points, this approximation is well suited. Then we can write (\ref{Hr}) in a block diagonal form which allows us to get an exact expression for the energy eigenvalues, given by

\begin{eqnarray}
E_{s,\eta}^\pm = \pm \sqrt{ \ls^2 \left(\varepsilon -s \eta \right)^2 + \left(\frac{\hbar v_F}{R}\right)^2 \left( \frac{\eta}{2} -\left(m + \frac{\Phi}{\Phi_0}\right) \right)^2 } \nnm
\label{Er0}
\end{eqnarray}
where $\varepsilon = \ez/\ec =\ez \ell / \ls$. $m$, $\eta(\pm 1)$ and $s(\pm 1)$ are the angular quantum number, valley and spin index respectively. The positive energy bands for either valley and spin is shown in Fig.~\ref{figsiband} using the parameters for silicene.

\begin{figure}[h]
\centering
\includegraphics[width=0.4\textwidth]{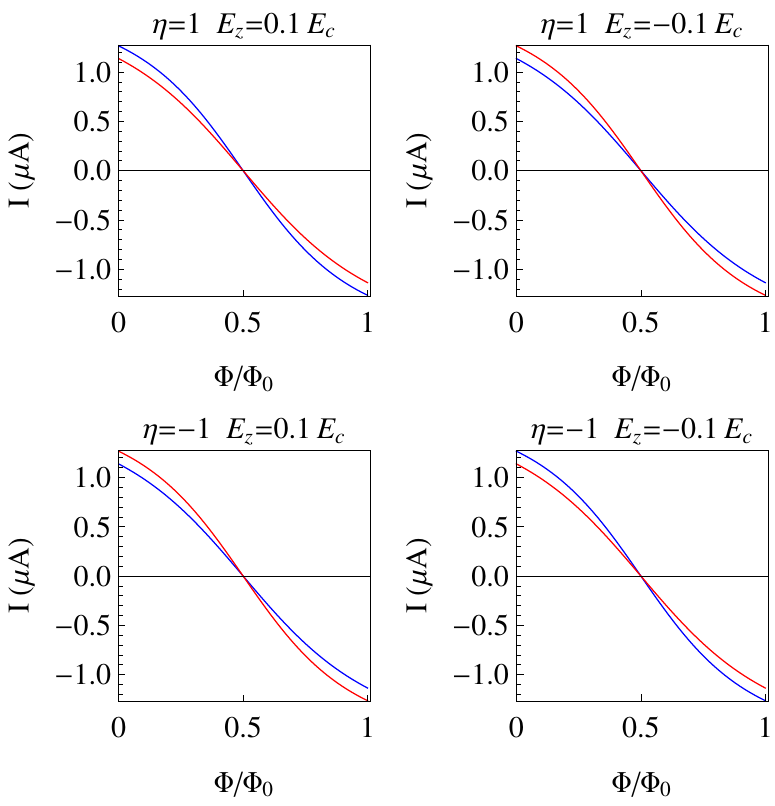}
\caption{Persistent current of silicene (Table\ref{param}) ring ($R=50\rm nm$) at $\ez=\pm 0.1 \ec$ and $\lr=0$ for . Blue and red lines corresponds spin up and spin down.}
\label{figsicur}
\end{figure}

As one can readily see band minima occur at half integral fluxes, which is a signature of relativistic particle \cite{Cotaescu2007a}. A gapless mode is formed when this condition is aided by $\varepsilon = s \eta$. 
The \pc\ in such a quantum ring is defined as \cite{Cheung1988} 

\begin{eqnarray}
I_{s,\eta} =- \Phi_0 \frac{e}{h} \frac{\partial E_{s,\eta}}{\partial \Phi}
\end{eqnarray}
and is shown in Fig.~\ref{figsicur} 

\begin{figure}[h]
\centering
\includegraphics[width=0.4\textwidth]{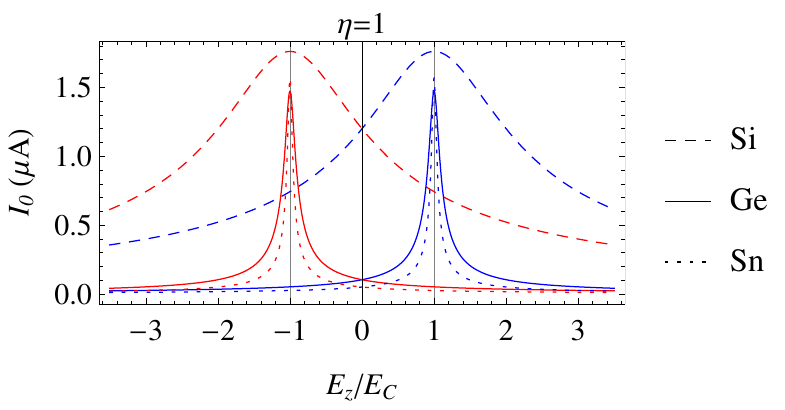}
\caption{ Maximum \pc\ of a ring ($R=50\rm nm$) for different $\ez$. Blue and red lines corresponds spin up and spin down.}
\label{figmax}
\end{figure}

As one can see, the effect of band gap is clearly manifested in the amplitude of the \pc . A smaller band gap gives higher curvature of the band resulting in a larger current. However they do not tell anything about the topological transition of the system. To determine the phase transition, let us examine the variation of maximum current with the applied electric field that occurs at $\Phi=0,\Phi_0$  (Fig.~\ref{figmax}). In this work we focus at $\Phi=0$.

\begin{figure}[h]
\centering
\includegraphics[width=0.4\textwidth]{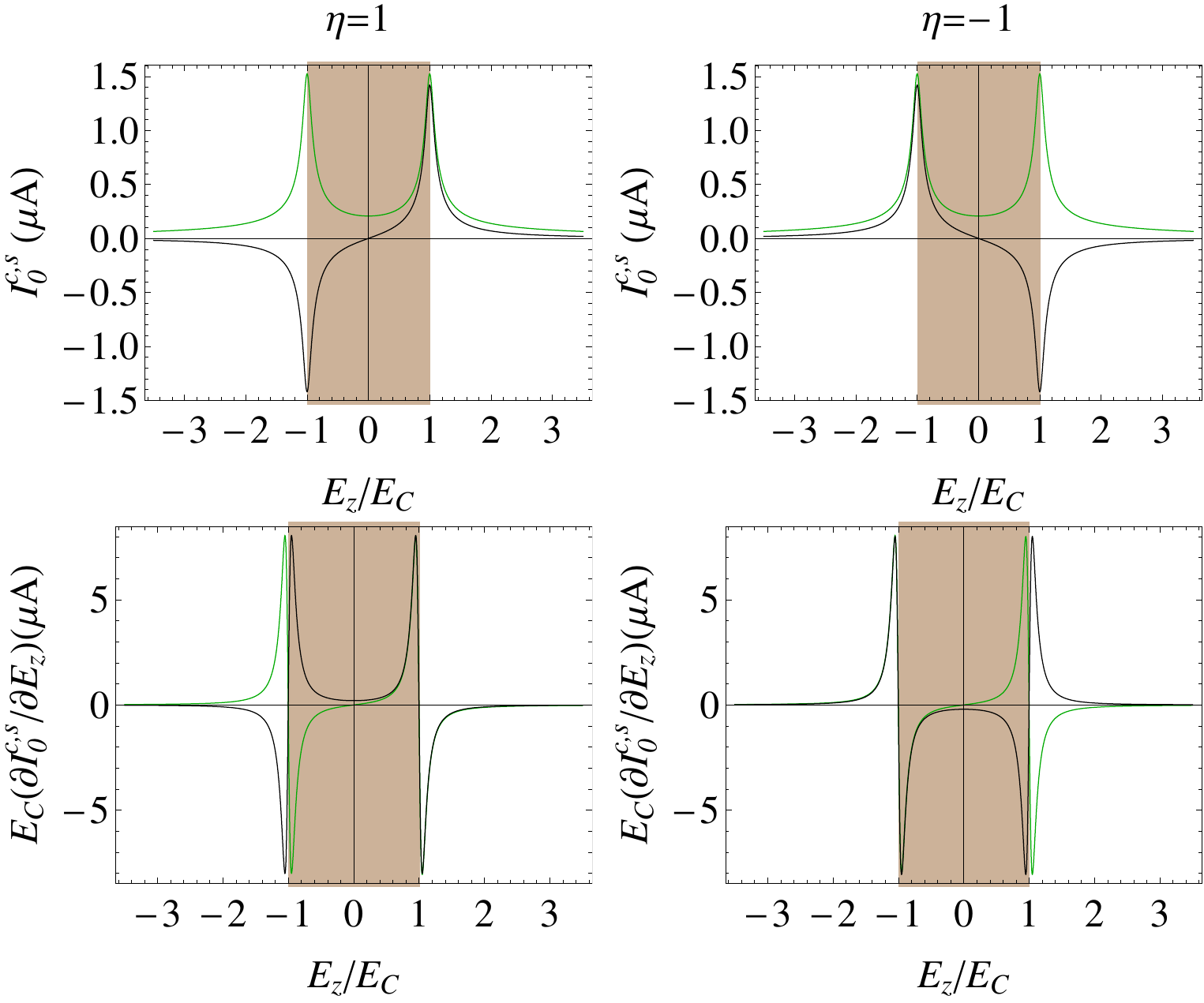}
\caption{ Variation of maximum charge (green) and spin (black) \pc\ and its derivative at $\Phi=0$ for silicene (Table\ref{param}) ring ($R=50\rm nm$) for different $\ez$. The brown box shows the topologically non trivial region.}
\label{figsichsp}
\end{figure}

\begin{figure}[h]
\centering
\includegraphics[width=0.25\textwidth]{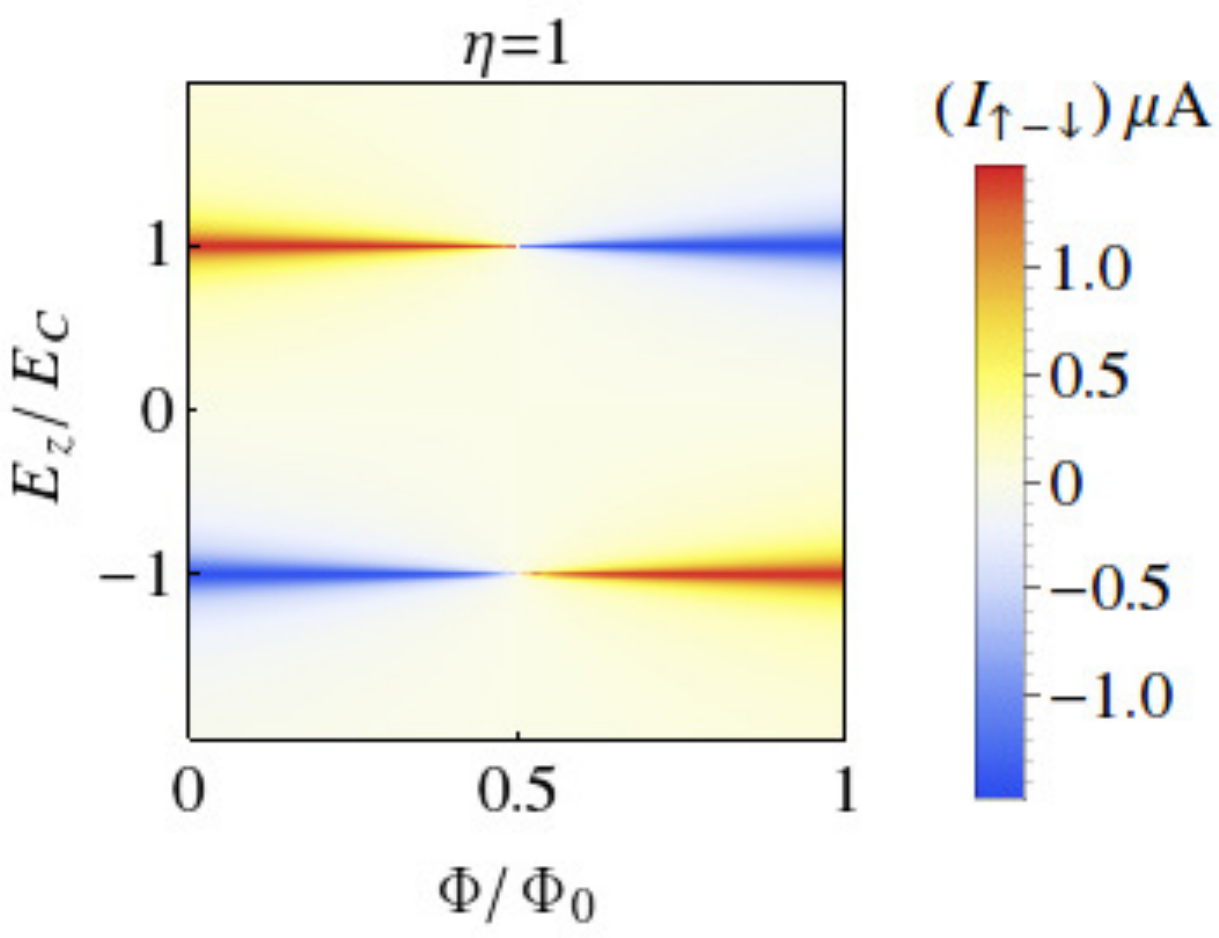}
\caption{Maximum spin current at different $\Phi$ and $\ez$ for a germanene ring ($R=50 \rm nm$).}
\label{figsiimax}
\end{figure}

The maximum current clearly denotes the point at which the transition is happening but it does not reveal anything about the topological phase. To extract that information we study the charge ($I_{0}^{c} = I_{0}^{\uparrow} + I_{0}^{\downarrow}$) and spin ($I_{0}^{c} = I_{0}^{\uparrow} - I_{0}^{\downarrow}$) current and its response to the change of $\ez$ (Fig.~\ref{figsichsp}). From now on we will use the parameters of germanene  (Fig.~\ref{figmax}).
From Fig.~\ref{figsichsp} one can clearly see that while the charge current remains positive and same for both valleys, the spin current changes sign with $\ez$ or with the valley index. The spin current can also reveal the topological phase if one examines its variation ($\partial I_0^s/\partial \ez$) with $\ez$. As we can see from Fig.~\ref{figsichsp}, the sign of $\partial I_0^s/\partial \ez$ clearly denotes the valley-spin-Chern number of the system and thus can be exploited to detect the topological phase of the system.
In principle this analysis of spin current can be done at any $\Phi$, but maximum contrast in spin current occurs near integer $\Phi/\Phi_0$ (Fig.~\ref{figsiimax}) which provides better identification.
  
\section{Effect of Rashba ($\lr$) coupling}

\begin{figure}[h]
\centering
\includegraphics[width=0.5\textwidth]{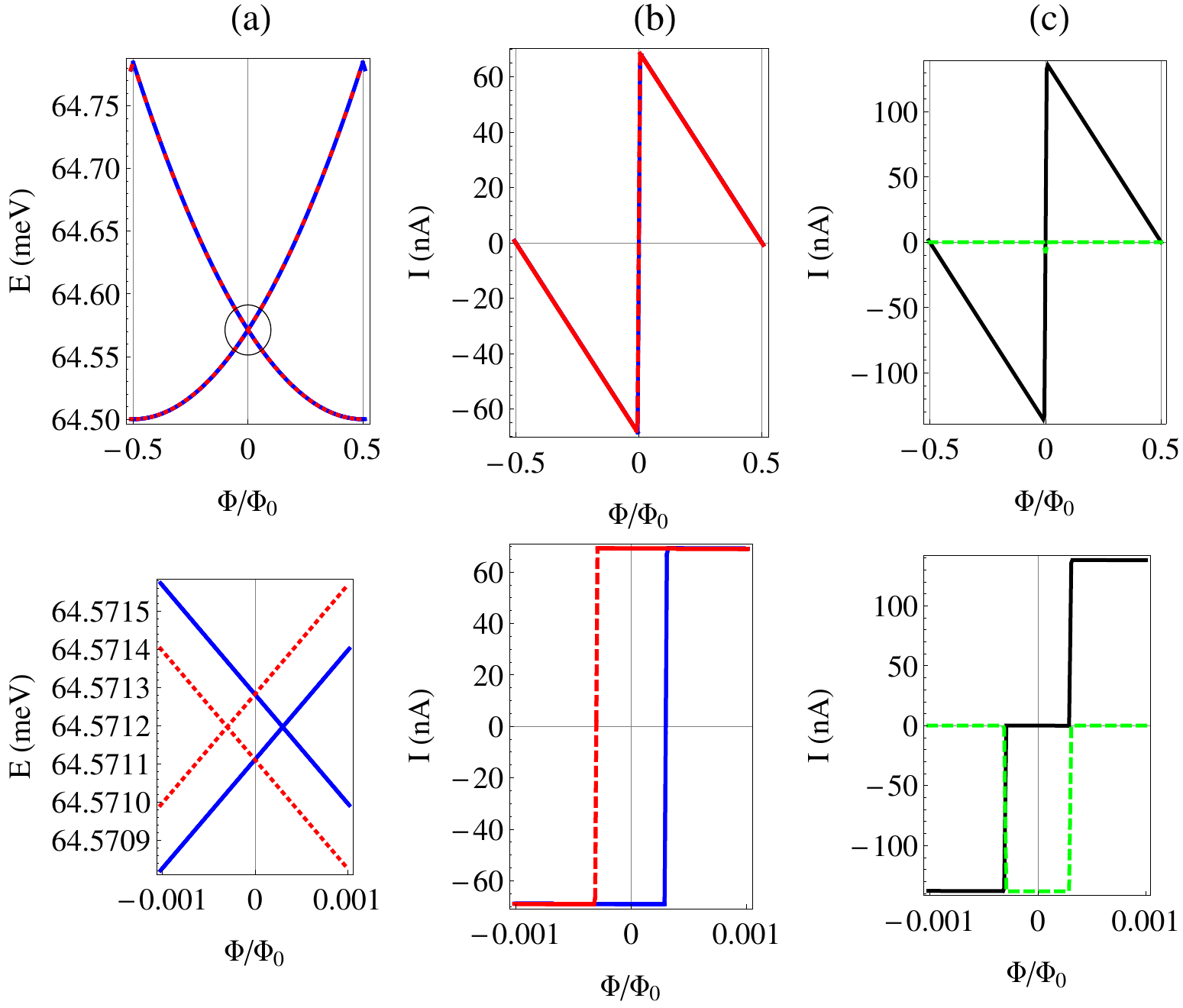}
\caption{(a) Energy dispersion, (b) \pc\ and (c) valley current for a germanene ring ($R=50\rm nm$) at $\ez=2.5\ec$. The bottom panel is enlarged version of the top panel neat $\Phi=0$. In (a) and (b) the solid and dashed line correspond $\eta=\pm 1$ and red and blue line corresponds spin up and down. In (c) the black and green dashed line correspond valley-charge ($I_0^{+}+I_0^{-}$) and valley-spin ($I_0^{+}-I_0^{-}$) current, the $\pm$ in the suffix denoting the valley.}
\label{figgeelr}
\end{figure}

From Table~\ref{param} we can see that Rashba coupling ($\lr$) is smaller than the spin orbit coupling ($\ls$) and hence has quite a negligible effect. For this reason we choose germanene where $\lr$ is comparatively stronger, to show the effect of Rashba spin-orbit coupling. In presence of Rashba spin-orbit coupling, we have to work with the full Hamiltonian (\ref{H}). An analytic solution is quite complicated and we proceed with numerical solutions. First we calculate the energy spectrum and the \pc\ for the system (Fig.~\ref{figgeelr}). In this section we are mostly interested in properties at zero flux so we keep our studies within the flux range $(-\Phi_0/2,\Phi_0/2)$. We also zoomed near $\Phi=0$ for better observation.

From Fig.~\ref{figgeelr} one can clearly see that a finite $\lr$ breaks the valley degeneracy for $\Phi\neq 0$.  This effect is completely different from the broken valley degeneracy that appears in a wide ring \cite{Recher2007} in absence of any spin dependent interaction. In case of a wide ring the effect is caused by the valley mixing due to infinite mass boundary condition and depends on the width of the ring. In our case it is coming from the momentum dependence of Rashba spin orbit coupling. From free electron model we know that Rashba spin orbit interaction acts like a gauge field which causes a shift in energy spectrum, keeping the degeneracy at zero momentum intact. In case of a buckled honeycomb lattice the effect of Rashba spin orbit coupling also depends on the valley index (\ref{H}). Besides the the presence of Aharonov-Bohm flux breaks the clockwise-anticlockwise symmetry. As a consequence for a particular valley the band minima shift only in a particular direction resulting a finite Aharonov-Casher phase. Same shift also takes place for the momentum at which the bands of same valley crosses each other (Fig.~\ref{figgeelr}a), which has opposite sign for opposite valley.  The maximum \pc\ occurs at the band crossing and hence it shifts from zero flux as well. In presence of a perturbation that couples the valleys, it results in a vanishing valley charge and finite valley spin current (Fig.~\ref{figgeelr}c) within a small flux window near $\Phi=0$. One should note that for $\lr=0$ the spin currents are opposite in either valleys (Fig.~\ref{figsichsp}) and cancel each other. Thus, a nonvanishing spin-valley current near zero flux is a signature of the Rashba spin orbit coupling. 

\begin{figure}[h]
\centering
\includegraphics[scale=0.75]{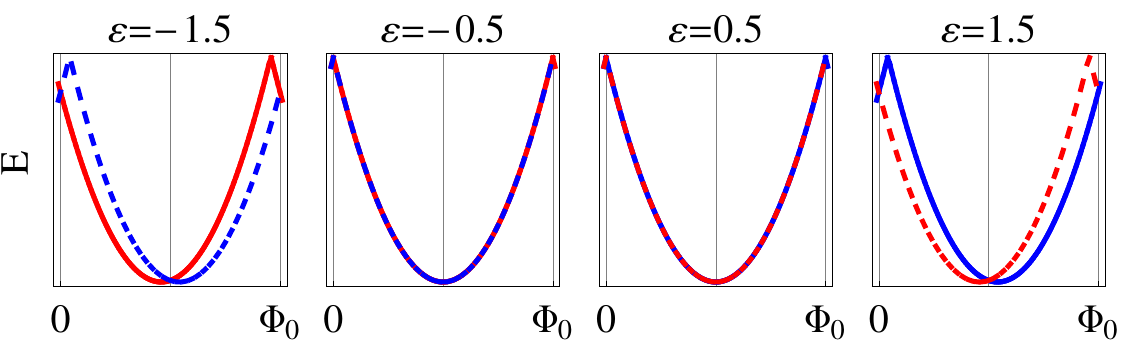}
\caption{Shift of energy spectrum due to Rashba spin orbit coupling in different topological regime for germanene ring. We increase the value of $\lr$ to make the effect visible.}
\label{rsocr}
\end{figure} 

\begin{figure}[h]
\centering
\includegraphics[scale=.75]{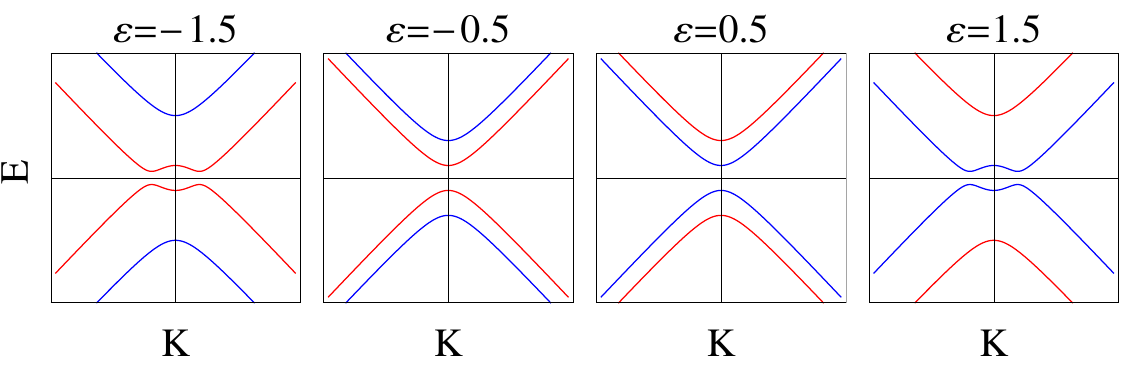}
\caption{Band structure of two dimensional honeycomb buckled lattice in presence of Rashba spin orbit coupling (in arbitrary units). $|\varepsilon|<1$ corresponds topologically nontrivial regime. Blue and red lines denotes spin up and spin down.}
\label{rsoc}
\end{figure}

Next we focus on the Aharonov-Casher phase in topologically nontrivial regime characterized by an applied electric field $\varepsilon = |\ez/\ec| <1$. First we calculate the energy spectrum at different field strength (Fig.~\ref{rsocr}).
From Fig.~\ref{rsocr} one can clearly see that the energy minimum is displaced from $\Phi_0/2$ only in topologically trivial regime. In bulk insulator regime the degeneracy is lifted everywhere except $\phi=0, \pm\Phi_0/2, \pm\Phi_0 \cdots$. For a qualitative understanding of this shift, we start from an infinite lattice (\ref{H}) where we can find an analytic expression for energy eigenvalues, given by

\begin{eqnarray}
E_{\eta,s}^\pm = \pm \sqrt{(\hbar v_F k)^2 + \ls^2 (\varepsilon +\eta s \sqrt{1+\alpha^2 k^2})^2 }
\label{elat}
\end{eqnarray}
where $\alpha = a \lr/\ls$. Equating $\partial E_{\eta,s}/ \partial k =0 $ gives the location of the maxima and minima of the spectrum. One can readily see that $k=0$ is an obvious solution. With a straight forward calculation one can easily show that the other two solutions can be real only if $\varepsilon^2 > (1+(\hbar v_F)^2/(\alpha \ls)^2)^2 $ (Fig.~\ref{rsoc}). Hence, the band minima at nonzero momentum appear only in topologically trivial regime. Or in other words, the gauge field due to Rashba spin-orbit coupling is real only in topologically trivial regime.  

\begin{figure}[h]
\centering
\includegraphics[width=0.25\textwidth]{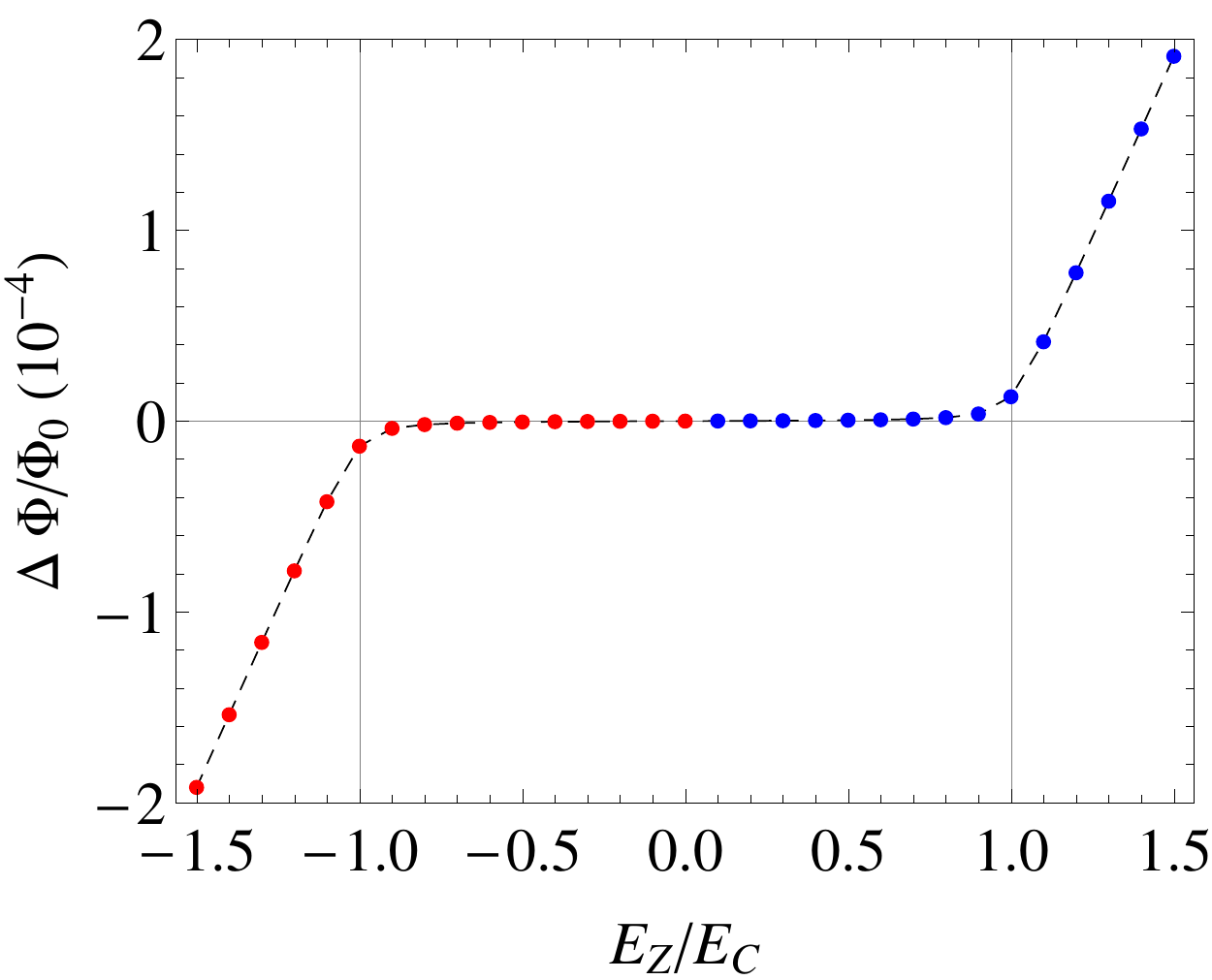}
\caption{Flux window for valley-spin current at different $\ez$. The color represents the ground state spin for $\eta=1$.}
\label{figgesplit}
\end{figure}

Similar effect is also present in a ring (Fig.~\ref{rsocr}), however the exact condition would be much complicated due to the involvement of radius of curvature and due to the fact that electrons with different spin or valley index accumulate different phases. 
From Fig.~\ref{rsocr} one can clearly see that the energy minima is displaced from $\Phi_0/2$ in topologically trivial regime which is analogous to the shift in a two dimensional lattice. The presence of the Aharonov-Bohm flux breaks the clockwise-anticlockwise symmetry, which gives a shift in only along one direction resulting a nonzero Aharonov-Casher phase. The Aharonov-Casher phase can be found from the intersection points of the bands that we calculate for different $\ez$ (Fig.~\ref{figgesplit}).

From Fig.~\ref{figgesplit} we can see that in the bulk insulator regime ($|\ez/\ec>1|$) increases monotonically. This is analogous to the band splitting we observe in a metallic or semiconducting ring due to spin orbit coupling. The flux window closes in the topologically nontrivial regime, which gives zero spin-valley current near zero flux in presence of a valley mixing perturbation. The spin-valley current thus can be utilized to detect the topological phase even in presence of valley mixing impurity.    

One should note that we introduce the gauge field through the substitution  $-i\frac{\partial}{\partial \theta} \rightarrow -i\frac{\partial}{\partial \theta} + \frac{\Phi}{\Phi_0}$. One can start with an opposite sign of the flux as well, which results in an opposite sign of the Aharonov-Casher phase. 


\section{Conclussion}
In this work we present a systematic study to identify different topological phases of a one dimensional Dirac ring by studying the persistent current. We show that the spin current and its derivative can distinctly identify the topological phase at each valley. We also consider the effect of Rashba interaction and show that the resulting Aharonov-Casher phase vanishes in topologically nontrivial regime. This results in a nonzero spin valley current for bulk insulators in presence of a valley mixing impurity, which can be utilized to detect the topological phases as well.

\section*{Acknowledgement}
S.G. would like to acknowledge helpful discussion with A. Saha and R. Tiwari. Research reported in this publication was supported by the King Abdullah University of Science and Technology (KAUST).

\bibliographystyle{apsrev4-1}

\bibliography{AB}

\begin{thebibliography}{28}%
\makeatletter
\providecommand \@ifxundefined [1]{%
 \@ifx{#1\undefined}
}%
\providecommand \@ifnum [1]{%
 \ifnum #1\expandafter \@firstoftwo
 \else \expandafter \@secondoftwo
 \fi
}%
\providecommand \@ifx [1]{%
 \ifx #1\expandafter \@firstoftwo
 \else \expandafter \@secondoftwo
 \fi
}%
\providecommand \natexlab [1]{#1}%
\providecommand \enquote  [1]{``#1''}%
\providecommand \bibnamefont  [1]{#1}%
\providecommand \bibfnamefont [1]{#1}%
\providecommand \citenamefont [1]{#1}%
\providecommand \href@noop [0]{\@secondoftwo}%
\providecommand \href [0]{\begingroup \@sanitize@url \@href}%
\providecommand \@href[1]{\@@startlink{#1}\@@href}%
\providecommand \@@href[1]{\endgroup#1\@@endlink}%
\providecommand \@sanitize@url [0]{\catcode `\\12\catcode `\$12\catcode
  `\&12\catcode `\#12\catcode `\^12\catcode `\_12\catcode `\%12\relax}%
\providecommand \@@startlink[1]{}%
\providecommand \@@endlink[0]{}%
\providecommand \url  [0]{\begingroup\@sanitize@url \@url }%
\providecommand \@url [1]{\endgroup\@href {#1}{\urlprefix }}%
\providecommand \urlprefix  [0]{URL }%
\providecommand \Eprint [0]{\href }%
\providecommand \doibase [0]{http://dx.doi.org/}%
\providecommand \selectlanguage [0]{\@gobble}%
\providecommand \bibinfo  [0]{\@secondoftwo}%
\providecommand \bibfield  [0]{\@secondoftwo}%
\providecommand \translation [1]{[#1]}%
\providecommand \BibitemOpen [0]{}%
\providecommand \bibitemStop [0]{}%
\providecommand \bibitemNoStop [0]{.\EOS\space}%
\providecommand \EOS [0]{\spacefactor3000\relax}%
\providecommand \BibitemShut  [1]{\csname bibitem#1\endcsname}%
\let\auto@bib@innerbib\@empty
\bibitem [{\citenamefont {B{\"{u}}ttiker}\ \emph {et~al.}(1983)\citenamefont
  {B{\"{u}}ttiker}, \citenamefont {Imry},\ and\ \citenamefont
  {Landauer}}]{Buttiker1983}%
  \BibitemOpen
  \bibfield  {author} {\bibinfo {author} {\bibfnamefont {M.}~\bibnamefont
  {B{\"{u}}ttiker}}, \bibinfo {author} {\bibfnamefont {Y.}~\bibnamefont
  {Imry}}, \ and\ \bibinfo {author} {\bibfnamefont {R.}~\bibnamefont
  {Landauer}},\ }\href {\doibase 10.1016/0375-9601(83)90011-7} {\bibfield
  {journal} {\bibinfo  {journal} {Phys. Lett. A}\ }\textbf {\bibinfo {volume}
  {96}},\ \bibinfo {pages} {365} (\bibinfo {year} {1983})}\BibitemShut
  {NoStop}%
\bibitem [{\citenamefont {Cheung}\ \emph {et~al.}(1988)\citenamefont {Cheung},
  \citenamefont {Gefen}, \citenamefont {Riedel},\ and\ \citenamefont
  {Shih}}]{Cheung1988}%
  \BibitemOpen
  \bibfield  {author} {\bibinfo {author} {\bibfnamefont {H.~F.}\ \bibnamefont
  {Cheung}}, \bibinfo {author} {\bibfnamefont {Y.}~\bibnamefont {Gefen}},
  \bibinfo {author} {\bibfnamefont {E.~K.}\ \bibnamefont {Riedel}}, \ and\
  \bibinfo {author} {\bibfnamefont {W.~H.}\ \bibnamefont {Shih}},\ }\href
  {\doibase 10.1103/PhysRevB.37.6050} {\bibfield  {journal} {\bibinfo
  {journal} {Phys. Rev. B}\ }\textbf {\bibinfo {volume} {37}},\ \bibinfo
  {pages} {6050} (\bibinfo {year} {1988})}\BibitemShut {NoStop}%
\bibitem [{\citenamefont {Loss}\ and\ \citenamefont
  {Goldbart}(1992)}]{Loss1992}%
  \BibitemOpen
  \bibfield  {author} {\bibinfo {author} {\bibfnamefont {D.}~\bibnamefont
  {Loss}}\ and\ \bibinfo {author} {\bibfnamefont {P.~M.}\ \bibnamefont
  {Goldbart}},\ }\href {\doibase 10.1103/PhysRevB.45.13544} {\bibfield
  {journal} {\bibinfo  {journal} {Phys. Rev. B}\ }\textbf {\bibinfo {volume}
  {45}},\ \bibinfo {pages} {13544} (\bibinfo {year} {1992})}\BibitemShut
  {NoStop}%
\bibitem [{\citenamefont {Chakraborty}\ and\ \citenamefont
  {Pietil\"{a}inen}(1995)}]{Chakraborty1995}%
  \BibitemOpen
  \bibfield  {author} {\bibinfo {author} {\bibfnamefont {T.}~\bibnamefont
  {Chakraborty}}\ and\ \bibinfo {author} {\bibfnamefont {P.}~\bibnamefont
  {Pietil\"{a}inen}},\ }\href {\doibase
  http://dx.doi.org/10.1103/PhysRevB.52.1932} {\bibfield  {journal} {\bibinfo
  {journal} {Phys. Rev. B}\ }\textbf {\bibinfo {volume} {52}},\ \bibinfo
  {pages} {1932} (\bibinfo {year} {1995})}\BibitemShut {NoStop}%
\bibitem [{\citenamefont {Levy}\ \emph {et~al.}(1990)\citenamefont {Levy},
  \citenamefont {Dolan}, \citenamefont {Dunsmuir},\ and\ \citenamefont
  {Bouchiat}}]{Levy1990}%
  \BibitemOpen
  \bibfield  {author} {\bibinfo {author} {\bibfnamefont {L.}~\bibnamefont
  {Levy}}, \bibinfo {author} {\bibfnamefont {G.}~\bibnamefont {Dolan}},
  \bibinfo {author} {\bibfnamefont {J.}~\bibnamefont {Dunsmuir}}, \ and\
  \bibinfo {author} {\bibfnamefont {H.}~\bibnamefont {Bouchiat}},\ }\href
  {http://journals.aps.org/prl/abstract/10.1103/PhysRevLett.64.2074} {\bibfield
   {journal} {\bibinfo  {journal} {Phys. Rev. Lett.}\ }\textbf {\bibinfo
  {volume} {64}},\ \bibinfo {pages} {2074} (\bibinfo {year}
  {1990})}\BibitemShut {NoStop}%
\bibitem [{\citenamefont {Mailly}\ \emph {et~al.}(1993)\citenamefont {Mailly},
  \citenamefont {Chapelier},\ and\ \citenamefont {Benoit}}]{Mailly1993}%
  \BibitemOpen
  \bibfield  {author} {\bibinfo {author} {\bibfnamefont {D.}~\bibnamefont
  {Mailly}}, \bibinfo {author} {\bibfnamefont {C.}~\bibnamefont {Chapelier}}, \
  and\ \bibinfo {author} {\bibfnamefont {A.}~\bibnamefont {Benoit}},\ }\href
  {http://journals.aps.org/prl/abstract/10.1103/PhysRevLett.70.2020} {\bibfield
   {journal} {\bibinfo  {journal} {Phys. Rev. Lett.}\ }\textbf {\bibinfo
  {volume} {70}} (\bibinfo {year} {1993})}\BibitemShut {NoStop}%
\bibitem [{\citenamefont {Jariwala}\ \emph {et~al.}(2001)\citenamefont
  {Jariwala}, \citenamefont {Mohanty}, \citenamefont {Ketchen},\ and\
  \citenamefont {Webb}}]{Jariwala2001}%
  \BibitemOpen
  \bibfield  {author} {\bibinfo {author} {\bibfnamefont {E.}~\bibnamefont
  {Jariwala}}, \bibinfo {author} {\bibfnamefont {P.}~\bibnamefont {Mohanty}},
  \bibinfo {author} {\bibfnamefont {M.}~\bibnamefont {Ketchen}}, \ and\
  \bibinfo {author} {\bibfnamefont {R.}~\bibnamefont {Webb}},\ }\href {\doibase
  10.1103/PhysRevLett.86.1594} {\bibfield  {journal} {\bibinfo  {journal}
  {Phys. Rev. Lett.}\ }\textbf {\bibinfo {volume} {86}},\ \bibinfo {pages}
  {1594} (\bibinfo {year} {2001})}\BibitemShut {NoStop}%
\bibitem [{\citenamefont {Bleszynski-Jayich}\ \emph {et~al.}(2009)\citenamefont
  {Bleszynski-Jayich}, \citenamefont {Shanks}, \citenamefont {Peaudecerf},
  \citenamefont {Ginossar}, \citenamefont {von Oppen}, \citenamefont
  {Glazman},\ and\ \citenamefont {Harris}}]{Bleszynski-Jayich2009}%
  \BibitemOpen
  \bibfield  {author} {\bibinfo {author} {\bibfnamefont {A.~C.}\ \bibnamefont
  {Bleszynski-Jayich}}, \bibinfo {author} {\bibfnamefont {W.~E.}\ \bibnamefont
  {Shanks}}, \bibinfo {author} {\bibfnamefont {B.}~\bibnamefont {Peaudecerf}},
  \bibinfo {author} {\bibfnamefont {E.}~\bibnamefont {Ginossar}}, \bibinfo
  {author} {\bibfnamefont {F.}~\bibnamefont {von Oppen}}, \bibinfo {author}
  {\bibfnamefont {L.}~\bibnamefont {Glazman}}, \ and\ \bibinfo {author}
  {\bibfnamefont {J.~G.~E.}\ \bibnamefont {Harris}},\ }\href {\doibase
  10.1126/science.1178139} {\bibfield  {journal} {\bibinfo  {journal}
  {Science}\ }\textbf {\bibinfo {volume} {326}},\ \bibinfo {pages} {272}
  (\bibinfo {year} {2009})}\BibitemShut {NoStop}%
\bibitem [{\citenamefont {Recher}\ \emph {et~al.}(2007)\citenamefont {Recher},
  \citenamefont {Trauzettel}, \citenamefont {Rycerz}, \citenamefont {Blanter},
  \citenamefont {Beenakker},\ and\ \citenamefont {Morpurgo}}]{Recher2007}%
  \BibitemOpen
  \bibfield  {author} {\bibinfo {author} {\bibfnamefont {P.}~\bibnamefont
  {Recher}}, \bibinfo {author} {\bibfnamefont {B.}~\bibnamefont {Trauzettel}},
  \bibinfo {author} {\bibfnamefont {A.}~\bibnamefont {Rycerz}}, \bibinfo
  {author} {\bibfnamefont {Y.}~\bibnamefont {Blanter}}, \bibinfo {author}
  {\bibfnamefont {C.}~\bibnamefont {Beenakker}}, \ and\ \bibinfo {author}
  {\bibfnamefont {A.}~\bibnamefont {Morpurgo}},\ }\href {\doibase
  10.1103/PhysRevB.76.235404} {\bibfield  {journal} {\bibinfo  {journal} {Phys.
  Rev. B}\ }\textbf {\bibinfo {volume} {76}},\ \bibinfo {pages} {235404}
  (\bibinfo {year} {2007})}\BibitemShut {NoStop}%
\bibitem [{\citenamefont {Zarenia}\ \emph {et~al.}(2010)\citenamefont
  {Zarenia}, \citenamefont {Pereira}, \citenamefont {Chaves}, \citenamefont
  {Peeters},\ and\ \citenamefont {Farias}}]{Zarenia2010}%
  \BibitemOpen
  \bibfield  {author} {\bibinfo {author} {\bibfnamefont {M.}~\bibnamefont
  {Zarenia}}, \bibinfo {author} {\bibfnamefont {J.~M.}\ \bibnamefont
  {Pereira}}, \bibinfo {author} {\bibfnamefont {A.}~\bibnamefont {Chaves}},
  \bibinfo {author} {\bibfnamefont {F.~M.}\ \bibnamefont {Peeters}}, \ and\
  \bibinfo {author} {\bibfnamefont {G.~A.}\ \bibnamefont {Farias}},\ }\href
  {\doibase 10.1103/PhysRevB.81.045431} {\bibfield  {journal} {\bibinfo
  {journal} {Phys. Rev. B}\ }\textbf {\bibinfo {volume} {81}},\ \bibinfo
  {pages} {045431} (\bibinfo {year} {2010})}\BibitemShut {NoStop}%
\bibitem [{\citenamefont {Schelter}\ \emph {et~al.}(2012)\citenamefont
  {Schelter}, \citenamefont {Recher},\ and\ \citenamefont
  {Trauzettel}}]{Schelter2012}%
  \BibitemOpen
  \bibfield  {author} {\bibinfo {author} {\bibfnamefont {J.}~\bibnamefont
  {Schelter}}, \bibinfo {author} {\bibfnamefont {P.}~\bibnamefont {Recher}}, \
  and\ \bibinfo {author} {\bibfnamefont {B.}~\bibnamefont {Trauzettel}},\
  }\href {\doibase 10.1016/j.ssc.2012.04.039} {\bibfield  {journal} {\bibinfo
  {journal} {Solid State Commun.}\ }\textbf {\bibinfo {volume} {152}},\
  \bibinfo {pages} {1411} (\bibinfo {year} {2012})}\BibitemShut {NoStop}%
\bibitem [{\citenamefont {Russo}\ \emph {et~al.}(2008)\citenamefont {Russo},
  \citenamefont {Oostinga}, \citenamefont {Wehenkel}, \citenamefont {Heersche},
  \citenamefont {Sobhani}, \citenamefont {Vandersypen},\ and\ \citenamefont
  {Morpurgo}}]{Russo2008}%
  \BibitemOpen
  \bibfield  {author} {\bibinfo {author} {\bibfnamefont {S.}~\bibnamefont
  {Russo}}, \bibinfo {author} {\bibfnamefont {J.}~\bibnamefont {Oostinga}},
  \bibinfo {author} {\bibfnamefont {D.}~\bibnamefont {Wehenkel}}, \bibinfo
  {author} {\bibfnamefont {H.}~\bibnamefont {Heersche}}, \bibinfo {author}
  {\bibfnamefont {S.}~\bibnamefont {Sobhani}}, \bibinfo {author} {\bibfnamefont
  {L.}~\bibnamefont {Vandersypen}}, \ and\ \bibinfo {author} {\bibfnamefont
  {A.}~\bibnamefont {Morpurgo}},\ }\href {\doibase 10.1103/PhysRevB.77.085413}
  {\bibfield  {journal} {\bibinfo  {journal} {Phys. Rev. B}\ }\textbf {\bibinfo
  {volume} {77}},\ \bibinfo {pages} {085413} (\bibinfo {year}
  {2008})}\BibitemShut {NoStop}%
\bibitem [{\citenamefont {Peng}\ \emph {et~al.}(2010)\citenamefont {Peng},
  \citenamefont {Lai}, \citenamefont {Kong}, \citenamefont {Meister},
  \citenamefont {Chen}, \citenamefont {Qi}, \citenamefont {Zhang},
  \citenamefont {Shen},\ and\ \citenamefont {Cui}}]{Peng2010}%
  \BibitemOpen
  \bibfield  {author} {\bibinfo {author} {\bibfnamefont {H.}~\bibnamefont
  {Peng}}, \bibinfo {author} {\bibfnamefont {K.}~\bibnamefont {Lai}}, \bibinfo
  {author} {\bibfnamefont {D.}~\bibnamefont {Kong}}, \bibinfo {author}
  {\bibfnamefont {S.}~\bibnamefont {Meister}}, \bibinfo {author} {\bibfnamefont
  {Y.}~\bibnamefont {Chen}}, \bibinfo {author} {\bibfnamefont {X.-L.}\
  \bibnamefont {Qi}}, \bibinfo {author} {\bibfnamefont {S.-C.}\ \bibnamefont
  {Zhang}}, \bibinfo {author} {\bibfnamefont {Z.-X.}\ \bibnamefont {Shen}}, \
  and\ \bibinfo {author} {\bibfnamefont {Y.}~\bibnamefont {Cui}},\ }\href
  {\doibase 10.1038/nmat2609} {\bibfield  {journal} {\bibinfo  {journal} {Nat.
  Mater.}\ }\textbf {\bibinfo {volume} {9}},\ \bibinfo {pages} {225} (\bibinfo
  {year} {2010})},\ \Eprint {http://arxiv.org/abs/0908.3314} {arXiv:0908.3314}
  \BibitemShut {NoStop}%
\bibitem [{\citenamefont {Fertig}\ and\ \citenamefont
  {Brey}(2010)}]{Fertig2010}%
  \BibitemOpen
  \bibfield  {author} {\bibinfo {author} {\bibfnamefont {H.~A.}\ \bibnamefont
  {Fertig}}\ and\ \bibinfo {author} {\bibfnamefont {L.}~\bibnamefont {Brey}},\
  }\href {\doibase 10.1098/rsta.2010.0210} {\bibfield  {journal} {\bibinfo
  {journal} {Philos. Trans. A. Math. Phys. Eng. Sci.}\ }\textbf {\bibinfo
  {volume} {368}},\ \bibinfo {pages} {5483} (\bibinfo {year}
  {2010})}\BibitemShut {NoStop}%
\bibitem [{\citenamefont {Yannouleas}\ \emph {et~al.}(2014)\citenamefont
  {Yannouleas}, \citenamefont {Romanovsky},\ and\ \citenamefont
  {Landman}}]{Yannouleas2014}%
  \BibitemOpen
  \bibfield  {author} {\bibinfo {author} {\bibfnamefont {C.}~\bibnamefont
  {Yannouleas}}, \bibinfo {author} {\bibfnamefont {I.}~\bibnamefont
  {Romanovsky}}, \ and\ \bibinfo {author} {\bibfnamefont {U.}~\bibnamefont
  {Landman}},\ }\href {\doibase 10.1103/PhysRevB.89.035432} {\bibfield
  {journal} {\bibinfo  {journal} {Phys. Rev. B}\ }\textbf {\bibinfo {volume}
  {89}},\ \bibinfo {pages} {035432} (\bibinfo {year} {2014})}\BibitemShut
  {NoStop}%
\bibitem [{\citenamefont {Kane}\ and\ \citenamefont {Mele}(2005)}]{Kane2005a}%
  \BibitemOpen
  \bibfield  {author} {\bibinfo {author} {\bibfnamefont {C.~L.}\ \bibnamefont
  {Kane}}\ and\ \bibinfo {author} {\bibfnamefont {E.~J.}\ \bibnamefont
  {Mele}},\ }\href {\doibase 10.1103/PhysRevLett.95.226801} {\bibfield
  {journal} {\bibinfo  {journal} {Phys. Rev. Lett.}\ }\textbf {\bibinfo
  {volume} {95}},\ \bibinfo {pages} {226801} (\bibinfo {year}
  {2005})}\BibitemShut {NoStop}%
\bibitem [{\citenamefont {Michetti}\ and\ \citenamefont
  {Recher}(2011)}]{Michetti2011}%
  \BibitemOpen
  \bibfield  {author} {\bibinfo {author} {\bibfnamefont {P.}~\bibnamefont
  {Michetti}}\ and\ \bibinfo {author} {\bibfnamefont {P.}~\bibnamefont
  {Recher}},\ }\href {\doibase 10.1103/PhysRevB.83.125420} {\bibfield
  {journal} {\bibinfo  {journal} {Phys. Rev. B}\ }\textbf {\bibinfo {volume}
  {83}},\ \bibinfo {pages} {125420} (\bibinfo {year} {2011})}\BibitemShut
  {NoStop}%
\bibitem [{\citenamefont {Zhou}\ \emph {et~al.}(2008)\citenamefont {Zhou},
  \citenamefont {Lu}, \citenamefont {Chu}, \citenamefont {Shen},\ and\
  \citenamefont {Niu}}]{Zhou2008}%
  \BibitemOpen
  \bibfield  {author} {\bibinfo {author} {\bibfnamefont {B.}~\bibnamefont
  {Zhou}}, \bibinfo {author} {\bibfnamefont {H.-Z.}\ \bibnamefont {Lu}},
  \bibinfo {author} {\bibfnamefont {R.-L.}\ \bibnamefont {Chu}}, \bibinfo
  {author} {\bibfnamefont {S.-Q.}\ \bibnamefont {Shen}}, \ and\ \bibinfo
  {author} {\bibfnamefont {Q.}~\bibnamefont {Niu}},\ }\href {\doibase
  10.1103/PhysRevLett.101.246807} {\bibfield  {journal} {\bibinfo  {journal}
  {Phys. Rev. Lett.}\ }\textbf {\bibinfo {volume} {101}},\ \bibinfo {pages}
  {246807} (\bibinfo {year} {2008})},\ \Eprint {http://arxiv.org/abs/0806.4810}
  {arXiv:0806.4810} \BibitemShut {NoStop}%
\bibitem [{\citenamefont {Ezawa}(2012{\natexlab{a}})}]{Ezawa2012a}%
  \BibitemOpen
  \bibfield  {author} {\bibinfo {author} {\bibfnamefont {M.}~\bibnamefont
  {Ezawa}},\ }\href {\doibase 10.1088/1367-2630/14/3/033003} {\bibfield
  {journal} {\bibinfo  {journal} {New J. Phys.}\ }\textbf {\bibinfo {volume}
  {14}},\ \bibinfo {pages} {033003} (\bibinfo {year}
  {2012}{\natexlab{a}})}\BibitemShut {NoStop}%
\bibitem [{\citenamefont {Drummond}\ \emph {et~al.}(2012)\citenamefont
  {Drummond}, \citenamefont {Z{\'{o}}lyomi},\ and\ \citenamefont
  {Fal'ko}}]{Drummond2012}%
  \BibitemOpen
  \bibfield  {author} {\bibinfo {author} {\bibfnamefont {N.~D.}\ \bibnamefont
  {Drummond}}, \bibinfo {author} {\bibfnamefont {V.}~\bibnamefont
  {Z{\'{o}}lyomi}}, \ and\ \bibinfo {author} {\bibfnamefont {V.~I.}\
  \bibnamefont {Fal'ko}},\ }\href {\doibase 10.1103/PhysRevB.85.075423}
  {\bibfield  {journal} {\bibinfo  {journal} {Phys. Rev. B}\ }\textbf {\bibinfo
  {volume} {85}},\ \bibinfo {pages} {075423} (\bibinfo {year}
  {2012})}\BibitemShut {NoStop}%
\bibitem [{\citenamefont {Ezawa}(2012{\natexlab{b}})}]{Ezawa2012}%
  \BibitemOpen
  \bibfield  {author} {\bibinfo {author} {\bibfnamefont {M.}~\bibnamefont
  {Ezawa}},\ }\href {\doibase 10.1103/PhysRevLett.109.055502} {\bibfield
  {journal} {\bibinfo  {journal} {Phys. Rev. Lett.}\ }\textbf {\bibinfo
  {volume} {109}},\ \bibinfo {pages} {055502} (\bibinfo {year}
  {2012}{\natexlab{b}})}\BibitemShut {NoStop}%
\bibitem [{\citenamefont {Liu}\ \emph {et~al.}(2011)\citenamefont {Liu},
  \citenamefont {Jiang},\ and\ \citenamefont {Yao}}]{Liu2011}%
  \BibitemOpen
  \bibfield  {author} {\bibinfo {author} {\bibfnamefont {C.-C.}\ \bibnamefont
  {Liu}}, \bibinfo {author} {\bibfnamefont {H.}~\bibnamefont {Jiang}}, \ and\
  \bibinfo {author} {\bibfnamefont {Y.}~\bibnamefont {Yao}},\ }\href {\doibase
  10.1103/PhysRevB.84.195430} {\bibfield  {journal} {\bibinfo  {journal} {Phys.
  Rev. B}\ }\textbf {\bibinfo {volume} {84}},\ \bibinfo {pages} {195430}
  (\bibinfo {year} {2011})}\BibitemShut {NoStop}%
\bibitem [{\citenamefont {Ezawa}(2015)}]{Ezawa2015a}%
  \BibitemOpen
  \bibfield  {author} {\bibinfo {author} {\bibfnamefont {M.}~\bibnamefont
  {Ezawa}},\ }\href {\doibase 10.7566/JPSJ.84.121003} {\bibfield  {journal}
  {\bibinfo  {journal} {J. Phys. Soc. Japan}\ }\textbf {\bibinfo {volume}
  {84}},\ \bibinfo {pages} {121003} (\bibinfo {year} {2015})},\ \Eprint
  {http://arxiv.org/abs/1503.0891} {arXiv:1503.0891} \BibitemShut {NoStop}%
\bibitem [{\citenamefont {Meijer}\ \emph {et~al.}(2002)\citenamefont {Meijer},
  \citenamefont {Morpurgo},\ and\ \citenamefont {Klapwijk}}]{Meijer2002}%
  \BibitemOpen
  \bibfield  {author} {\bibinfo {author} {\bibfnamefont {F.}~\bibnamefont
  {Meijer}}, \bibinfo {author} {\bibfnamefont {A.}~\bibnamefont {Morpurgo}}, \
  and\ \bibinfo {author} {\bibfnamefont {T.}~\bibnamefont {Klapwijk}},\ }\href
  {\doibase 10.1103/PhysRevB.66.033107} {\bibfield  {journal} {\bibinfo
  {journal} {Phys. Rev. B}\ }\textbf {\bibinfo {volume} {66}},\ \bibinfo
  {pages} {033107} (\bibinfo {year} {2002})}\BibitemShut {NoStop}%
\bibitem [{\citenamefont {Wang}\ and\ \citenamefont
  {Manchon}(2011)}]{Wang2011}%
  \BibitemOpen
  \bibfield  {author} {\bibinfo {author} {\bibfnamefont {X.}~\bibnamefont
  {Wang}}\ and\ \bibinfo {author} {\bibfnamefont {A.}~\bibnamefont {Manchon}},\
  }\href {\doibase 10.1063/1.3647569} {\bibfield  {journal} {\bibinfo
  {journal} {Appl. Phys. Lett.}\ }\textbf {\bibinfo {volume} {99}},\ \bibinfo
  {pages} {142507} (\bibinfo {year} {2011})},\ \Eprint
  {http://arxiv.org/abs/1111.1244} {arXiv:1111.1244} \BibitemShut {NoStop}%
\bibitem [{\citenamefont {Ghosh}\ and\ \citenamefont {Saha}(2014)}]{Ghosh2014}%
  \BibitemOpen
  \bibfield  {author} {\bibinfo {author} {\bibfnamefont {S.}~\bibnamefont
  {Ghosh}}\ and\ \bibinfo {author} {\bibfnamefont {A.}~\bibnamefont {Saha}},\
  }\href {\doibase 10.1140/epjb/e2014-50223-1} {\bibfield  {journal} {\bibinfo
  {journal} {Eur. Phys. J. B}\ }\textbf {\bibinfo {volume} {87}},\ \bibinfo
  {pages} {167} (\bibinfo {year} {2014})}\BibitemShut {NoStop}%
\bibitem [{\citenamefont {Ghosh}(2013)}]{Ghosh2013}%
  \BibitemOpen
  \bibfield  {author} {\bibinfo {author} {\bibfnamefont {S.}~\bibnamefont
  {Ghosh}},\ }\href {\doibase 10.1155/2013/592402} {\bibfield  {journal}
  {\bibinfo  {journal} {Adv. Condens. Matter Phys.}\ }\textbf {\bibinfo
  {volume} {2013}},\ \bibinfo {pages} {592402} (\bibinfo {year}
  {2013})}\BibitemShut {NoStop}%
\bibitem [{\citenamefont {Cotaescu}\ and\ \citenamefont
  {Papp}(2007)}]{Cotaescu2007a}%
  \BibitemOpen
  \bibfield  {author} {\bibinfo {author} {\bibfnamefont {I.~I.}\ \bibnamefont
  {Cotaescu}}\ and\ \bibinfo {author} {\bibfnamefont {E.}~\bibnamefont
  {Papp}},\ }\href {\doibase 10.1088/0953-8984/19/24/242206} {\bibfield
  {journal} {\bibinfo  {journal} {J. Phys. Condens. Matter}\ }\textbf {\bibinfo
  {volume} {19}},\ \bibinfo {pages} {242206} (\bibinfo {year}
  {2007})}\BibitemShut {NoStop}%
\end{thebibliography}%
\end{document}